\documentclass[printer]{aa} 
\usepackage{graphicx}
\usepackage{txfonts}
\newcommand{\Y}{Y_\ell^m}

\newcommand{\bee}{\begin{eqnarray}}
\newcommand{\ene}{\end{eqnarray}}
\newcommand{\pol}{{1\over2}}
\newcommand{\ve}{\mbox{\boldmath{$e$}}}
\newcommand{\rd}{\rm d}
\newcommand{\vnab}{\mbox{\boldmath{$\nabla$}}}

\newcommand{\vx}{\mbox{\boldmath{$x$}}}

\newcommand{\vxi}{\mbox{\boldmath{$\xi$}}}

\begin{document}
   \title{Dipolar Modes in Luminous Red Giants}
   \author{W. A. Dziembowski}
\institute{Warsaw University Observatory,
              Aleje Ujazdowskie 4, 00-478\\
             and Copernicus Astronomical Center, ul. Bartycka 18, 00-787
             Warsaw, Poland\\
              \email {wd@astrouw.edu.pl}\\}
   \date{Received July 20, 2011; accepted January 6, 2012}


  \abstract
   {Lots of information on solar-like oscillations in red giants has been obtained thanks to observations
   with CoRoT and \it Kepler \rm space telescopes. Data on dipolar modes appear most interesting.}
   {We study properties of dipolar oscillations in luminous red giant to explain mechanism of
   mode trapping in the convective envelope and to assess what may be learned from the new data.}
   {Equations for adiabatic oscillations are solved by numerical integration down to the bottom
   of convective envelope, where the boundary condition is applied. The condition is based on
   asymptotic  decomposition of the fourth order system into components describing a running wave
   and a nearly unform shift of radiative core.}
   {If the luminosity of a red giant is sufficiently high ($L\gtrsim 100L_\odot$ at $M=2M_\odot$),
   the dipolar modes become effectively trapped in the acoustic cavity, which
   covers the outer part of convective envelope. Energy loss caused by gravity wave emission at the envelope base is a
   secondary or negligible source of damping. Frequencies are insensitive to structure of the deep interior.}
   {}


   \maketitle
\section{Introduction}

Observations with the space-borne telescopes CoRoT and Kepler
brought wealth of information about solar-like oscillations in stars
along the red giants branch (RGB) (Mosser et al. 2011, Hekker et
al., 2011, and references therein). In his recent review,
Christensen-Dalsgaard (2011) describes these objects as "the most
interesting areas of space asteroseismology". Here we focus on
dipolar modes in these stars not only because such modes are most
important for stellar, in particular red giants (see
Montalb\'an  et al. 2011), diagnostic but also because they present some
interesting problems for stellar pulsation theory.

What makes dipolar oscillations different from those of higher
degrees is that the Cowling approximation is not applicable. This
approximation has been the basis of standard mode classification and
analysis of wave propagation. The problem of a unique dipolar mode
classification was solved by Takata (2005, 2006b). The aspect of our
interest in this paper is the effect of the wave trapping in the
acoustic cavity which is essential for understanding oscillations
observed in red giants.

The gross picture of red giant oscillation emerging from analysis of
massive data on red giant oscillations (Mosser et al. 2010)  turns out rather
simple.   The power spectra look similar to that of the Sun from spatially unresolved  measurements.
The sequences of low degree modes, up to $\ell=3$, are very clearly seen.
Only after a closer look a difference in
 dipolar mode pattern may be noticed. This difference is important and interesting.

Observations largely confirmed what has been predicted by
theoretical studies of Dupret et al. (2009) and Montalb\'an  et al. (2010).
The first paper was devoted to stochastic excitation of oscillation in red giants and
gives values of expected amplitudes and lifetimes for five selected
red giants models. The second paper presented an extensive survey of
the trapped mode frequencies in red giant models. In very dense
eigenfrequency spectra, the modes were identified as the ones corresponding to local minima of
inertia. Such choice of the  modes was motivated by the results of the first paper and
two earlier papers (Dziembowski et al, 2001, Christensen-Dalsgaard, 2004),
where importance of mode trapping phenomenon in red giant oscillations
was stressed. The goal of the present work is to improve understanding of
 this phenomenon and its consequences for red giant seismology.

The next section is devoted to derivation of an asymptotic approximation for
dipolar oscillations in the gravity propagation zone and a discussion
of  transition from standing to running waves limit. Based on that, in Sect. 3,
the inner boundary condition for the acoustic oscillation of the convective envelope is formulated.
Properties of the acoustic cavity in red giant envelopes
are reviewed in Sect. 4. Efficiency of trapping radial and dipolar
modes is discussed in Sect. 5. Numerical results on mode frequencies and damping rates caused by wave
losses in selected envelope models are presented, respectively, in Sect. 6 and 7.
The final section confronts theoretical results obtained in this and earlier
papers with observations.

\section{An asymptotic approximation for dipolar modes}
We consider adiabatic oscillation of an isolated star ignoring all effects
of rotation.
The linearized  momentum equation, which we write in the standard notation adopted
e.g. by Aerts et al. (2010)
\begin{equation}
\omega^2\rho\vxi=\vnab p'+\rho\vnab\Phi'+\rho' g\ve_r,
\end{equation}
combined with the Poisson and continuity equations leads to the
familiar fourth order differential equations for the radial
eigenfunctions.

That the system reduces to the second order at $\ell=0$ (radial modes) has been known from
the nascent of the theory of stellar oscillations but that this is true  also at $\ell=1$
(dipolar modes) has been demonstrated not long ago
by Takata (2005). Here we we will use his formalism and mostly his notation.
The four basic radial eigenfunctions $y_k(r)$ used in this work are defined in
expressions for vectorial displacement
\begin{equation}
\vxi =r[y_1(r)\ve_r+y_2(r)\vnab_H]\Y {\rm e}^{-{\rm i}\omega},
\end{equation}
and the Eulerian perturbation of the gravitational potential
\begin{equation}
\vnab\Phi'=g[y_4(r)\ve_r+y_3(r)\vnab_H]\Y{\rm e}^{-{\rm i}\omega t}.
\end{equation}
Note that we have
$$
 p'=g\rho r[By_2(r)-y_3(r)]\Y{\rm e}^{-{\rm
i}\omega t}, $$
 where $$B={\omega^2r\over g}.$$ We will use standard
dimensionless coefficients calculated from stellar models
$$V_g={gr\over c^2}, \qquad A=-{d\ln\rho\over d\ln r}-V_g, \quad {\rm
and}\quad U={4\pi\rho r^3\over M_r}.$$
Takata showed that the radial eigenfunctions of dipolar modes
satisfy the linear algebraic relation
\begin{equation}
UB(y_2-y_1)+(2-U+B)y_3+(1-B)y_4=0
\end{equation}
and used it to reduce the eigenvalue problem to two first order
equations for radial eigenfunctions
\begin{equation}
z_1=y_1+{y_3\over1-B}
\end{equation}
and
\begin{equation}
z_2=y_2+{y_3\over1-B}.
\end{equation}
Here, we write these equations in the following form
\begin{equation}
r{\rd\over\rd r}\left(\matrix{z_1\cr
z_2}\right)=\left(\matrix{V_g-3+S&2-BV_g-S\cr
1-A/B+S&A-2-S}\right)\left(\matrix{z_1\cr z_2}\right),
\end{equation}
where
$$S={UB\over(1-B)^2}.$$

For the radial eigenfunctions describing perturbation we have from
the definition in Eq. 3
\begin{equation}
r{\rd y_3\over\rd r}=(1-U)y_3+y_4
\end{equation}
and from Eq. 4
\begin{equation}
y_4={(U-2-B)y_3+UB(z_1-z_2)\over{1-B}}.
\end{equation}
At the base of red giant envelope, $B\propto r^3/M_r$ may be very
small even for high frequency modes. Let us note that  Eq. 7 in the
$B\rightarrow0$ limit has the same form as in the Cowling
approximation but with modified definition of variables. Nearly the
whole radiative core, where $A>0$, is a region of gravity wave
propagation. Once the wave number, $\sqrt{2A/Br^2}$, is
sufficiently large an asymptotic approximation is applicable.

For any $\ell>0$ in the limit $A/B\gg1$, the fourth order system may
be  asymptotically decomposed into two second order equations
yielding one solution that varies rapidly with $r$ and describes a
gravity wave and one that varies slowly and describes a quasi-static
response to an external perturbation of the gravitational potential
(Dziembowski, 1971).  In 1971, the goal of such a decomposition was
not only to help understanding physics of nonradial oscillations in
evolved stars but also to use in the interior the analytic form of
the rapidly varying solution which was difficult to calculate
numerically.

In that work no special attention was paid to dipolar modes for
which also the slow solution may be derived in an analytical form.
With Takata's equations, the solution for the potential which is
divergent at $r=0$ is eliminated automatically.  A superposition of
the slow and rapid solutions will be used in the next section to
formulate inner boundary conditions for dipolar modes trapped in the
convective envelope.

The slowly varying solution of Eqs. 7 to 9, which is easily
obtained by setting $z_1=z_2=0$, is
\begin{equation}
\left(\matrix{y_{s,1}\cr y_{s,2}\cr y_{s,3}\cr
y_{s,4}}\right)={C_s\over r}
\left(\matrix{1\cr1\cr-1+B\cr2-U+B}\right),
\end{equation}
where $C_s$ is a constant. This solution is exact and it is
equivalent to one given by Takata (2006a, Eq. 116). It describes
uniform shift of star interior in response to the external potential
change, which in our application is caused by oscillations in the
envelope.

To obtain the rapidly varying solution, we adopt
$$\left\vert{\rd z_k\over\rd r}\right\vert\sim |k_r z_k|,$$
where
\begin{equation}
k_r={1\over r}\sqrt{2A\over B}\gg{1\over r}
\end{equation}
is the radial component of gravity wave number. Setting in Eq. 7
\begin{equation}
z_1={{\cal Z}_1\over r^3\sqrt\rho},
\end{equation}
we obtain
\begin{equation}
r^2{\rd^2{\cal Z}_1\over\rd r^2}+\left[{2A\over B} +{\cal O}(1)\right]{\cal Z}_1=0.
\end{equation}
The asymptotic solution of this equation is
\begin{equation}
{\cal Z}_1=C_r{{\rm e}^{{\rm i}\Psi}+{\rm e}^{-{\rm i}\Psi}\over\sqrt{k_r}},
\quad  \Psi=\int_{r_0} k_r dr,
\end{equation}
where $C_r$ is another constant. If there is no convective core
$r_0=0$ and this form is obtained by fitting to the spherical Bessel
function $j_1(r/R)$, which is the solution of Eq. 7 valid near the
center. The same form is valid for stars with convective cores but
then $r_0=r_c$, which is the core radius. The first and second
exponential components describe, respectively, outwardly and
inwardly running gravity waves. At $B\ll1$,  Eqs. 7 and 14 imply
\begin{equation}
z_2\approx{1\over 2}\left( r{\rd z_1\over\rd r}+(3-V_g)z_1\right)={{\cal Z}_2\over 2r^3\sqrt\rho},
\end{equation}
where
\begin{equation}
{\cal Z}_2=r{\rd{\cal  Z}_1\over\rd r}+{A-V_g\over 2}{\cal Z}_1.
\end{equation}
With the same accuracy,  Eqs. 5 and 6 imply $y_1=z_1$ and $y_2=z_2$.
Then, Eqs. 8 and 9 yield $y_3=-UBy_1/2$ and
$y_4=(U-2)y_3+UB(z_1-z_2)$. Thus, for the rapidly varying solution,
we get the following asymptotic approximation,
\begin{equation}
 \left(\matrix{y_{r,1}\cr y_{r,2}\cr y_{r,3}\cr y_{r,4}}\right)=
 {C_r\over\sqrt{\rho} r^3}
\left(\matrix{2{\cal Z}_1\cr {\cal Z}_2\cr
-UB{\cal Z}_1\cr -UB[{\cal Z}_2+(U-4){\cal Z}_1]}\right).
\end{equation}

If energy losses in the interior are negligible, the
amplitudes of the inwardly and outwardly running waves are equal and
the solution is a standing wave. In luminous red giants even at
frequencies as high as the acoustic cut-off the phase at the top
gravity propagation zone is large. Thus, all nonradial modes have
mixed character. The distance between consecutive  modes of the same
$\ell\geq1$ is much less than the distance between consecutive
radial modes (the large frequency separation) and decreases during
evolution along the red giant branch. The relative amplitude in the
interior is modulated in the rhythm determined by the acoustic
frequencies of the envelope. With increasing luminosity, the
amplitude of modulation increases which leads to the appearance of
modes trapped in the envelope. This happens first for modes of
higher degrees but ultimately also for $\ell=1$. Such modes are of
our interest.

The most important evolutionary effect on nonadiabatic properties of
nonradial oscillations is the onset of significant damping in the inner part
of the radiative interior (Dziembowski, 1971).
In the first works on excitation of nonradial modes in giants
(Dziembowski 1977, Osaki 1977), only unfitted envelope models were
considered. Solutions of the nonadiabatic problem were obtained
upon neglecting perturbation of the gravitational
potential (the Cowling approximation). One needed
boundary condition was obtained by neglecting the reflected wave.
The justification was that radiative damping in the deep interior causes
that the reflected wave returns with much reduced amplitude and may be
ignored in some cases, such as low order modes of moderate
degrees in Cepheids and RR Lyrae stars considered in those papers.
The modes were found unstable in spite of large damping in the interior because
the amplitudes there were very low as a result of efficient trapping in the envelope.
For luminous red giants, such a situation was found also in the case
of dipolar modes (Dziembowski et al. 2001).

Complex eigenfrequencies determined with the running wave boundary
condition set in the envelope yield approximate values for true
stellar eigenfrequencies if $|\Psi_I(r_b)|\gg1$, where subscript $I$
denotes the imaginary part of a complex quantity (we will use
subscript $R$ to denote the real part) not only for unstable modes
but also for stochastically driven stable modes. The requirement is
that the damping rate is not too high. The explicit condition is
easy to write in the case when the quasi-adiabatic approximation is
still applicable and $|\omega_I|\ll\omega_R$. In such a special but
not unrealistic case we have (e.g. Dziembowski et al. 2001)
$$k_r\simeq{\sqrt{\ell(\ell+1)}\over\omega_R}{N\over r}[1-{\rm i}(D+{\omega_I\over\omega_R})],$$
where
$$ D={\ell(\ell+1)\over 8\pi\omega^3} {gL_{\rm r}\over r^4p}
         {\nabla_{\rm ad}\over\nabla}(\nabla_{\rm ad}-\nabla),$$
and this condition is
\begin{equation}
-\Psi_I(r_b)=\int_0^{\Psi_b}\left(D -{\omega_I\over\omega_R}\right)\rd\Psi_R\gg1,
\end{equation}
where $\Psi_b=\Psi_R(r_b)$.
The trapping effect follows essentially from adiabatic properties of stellar oscillation but it is
enhanced by nonadiabatic effects in the case of unstable or weakly stable modes.

 As long as we stay within the domain of linear oscillation
theory, the use of running wave solution in the situation when the
strong inequality in Eq. 18 is not satisfied means an overestimate
of damping in the core. However, it must be kept in mind that
nonlinear effects may significantly enhance radiative damping. Could
such effects be of any importance in the case of stochastically
driven oscillations? Dupret et al. (2009) found that everywhere in
their models $|(\xi\cdot\vnab)\xi|\ll|\xi|$ and gave the negative
answer. However, their argument refers to a strong nonlinearity,
whereas at much lower amplitudes the resonant excitation of high
degree modes may occur which leads to a significant increase of the
damping rates (e.g. Kumar \& Goodman, 1996). Unfortunately, in this
case, a precise determination of the condition for the running wave
approximation is not easy.

In this paper, we adopt the running wave boundary condition at the
top of radiative interior. This means that we treat the convective
envelope as a separate oscillating body loosing part of its energy
by emitting gravity wave from its bottom. We will examine
implication of this assumption and appeal to observations to verify
its applicability.

\section{The boundary conditions and eigenfrequencies}

The asymptotic approximation developed in the previous section allows to formulate the inner boundary condition
which may be imposed on numerical solutions of the equations for dipolar oscillation in the outer layers.
Radiative dissipation arises mainly deep in the star interior, so that even when it is large,
we may still rely on adiabatic approximation in the outer part of the g-mode propagation zone,
where we want to apply the inner boundary condition. This is not an essential approximation and in fact
it was not used in Dziembowski (1977) paper.  However, we will avoid
the Cowling approximation, which  may be globally justified at higher degrees but not,
 as Christensen-Dalsgaard \& Gough (2001) stressed, at $\ell=1$.
We are now going to correct this mistake by taking into account perturbation of the gravitational
potential and the slowly varying solution.

The general solution of Eqs. 7 to 9 is a linear superposition of the
slow and rapid solutions given in Eqs. 10 and 17,
 respectively. It involves two unknown constants and thus yields two independent boundary conditions.
The slow solution has the form specific for dipolar modes and only for these modes it is independent
of the interior structure.
 The form of the rapid solution is valid for any $\ell$ if  in Eq. 11 for $k_r$
the number 2 is replaced with $\ell(\ell+1)$. For the running wave
solution, we get from Eqs. 14 and 16
\begin{equation}
{\cal Z}_2=\left[{\rm i}rk_r +\pol\left(A-V_g+{r\over|k_r|}{\rd|k_r|\over\rd r}\right)\right]
{\cal Z}_1.
\end{equation}
 The boundary condition obtained in this way may be applied at any place
within the radiative interior, where the asymptotic approximation is valid.
The explicit form of the inner boundary condition for the case of running wave is given later
in this section. The same conditions may be applied at the base
of convective envelope, $ r=r_b$, if  $B(r_b)\ll1$ and there is a jump in $A$ to a finite value due to
discontinuity in hydrogen abundance.
Note that then a corresponding discontinuity of $y_2$ must be taken into account.
However, it is more realistic to treat $A(r)$ as a steep but continuous function. In such a
case, some modifications in the form of boundary are needed.

To this end, we consider thin layer beneath the envelope bottom
where $A$ rises from zero at $r=r_b$ to a finite value at its lower
boundary where $A/B$ becomes large. Like before, we start with Eq.
13 but now we seek its asymptotic solution in the $B\rightarrow0$
limit in terms of the Airy function ${\cal A}i$ and ${\cal B}i$ (see
e.g.  Eq. 10.4.112 in Abramovitz\& Stegun  1972). The solution is
\begin{equation}
{\cal Z}_1=C_A{\cal A}i(s)+ C_B{\cal B}i(s).
\end{equation}
where
$$s=-\lambda^{2/3}(r_b-r),\quad \lambda^2=-
\left({2\over r^2B}{\rd A\over \rd r}\right)_b>0.$$
Note that
$$r{\rd {\cal Z}_1\over\rd r} =\beta{\rd {\cal Z}_1\over\rd s},$$
where
\begin{equation}
\beta\equiv r\lambda^{2/3}=\left(-{2r\over B}{\rd A\over\rd r}\right)^{1/3}_b.
\end{equation}

At $(r_b-r)\kappa\gg r_b$,
$${\cal Z}_1\propto |s|^{-1/4}(C_A\sin\Psi+C_B\cos\Psi),
\quad\Psi={2\over3}(-s)^{3\over2}+{\pi\over4}.$$
Thus, for the inwardly  running wave we must have $C_A=-{\rm i}C_r$, $C_r\equiv C_B$, and
we get
\begin{equation}
{\cal Z}_1={\cal B}i(s)-{\rm i}{\cal A}i(s).
\end{equation}
The expression for $z_2$ in Eq. 15 remains valid but with
\begin{equation}
{\cal Z}_2=\beta{\rd\over\rd s}({\cal B}i-{\rm i}{\cal A}i)+{A-V_g\over2}{\cal Z}_1
\end{equation}
Again  with Eqs. 8 and 9, we may derive the corresponding
expressions for $y_3$ and $y_4$. The slow solution given by Eq. (10)
remains unchanged. For the explicit form of the solution at
convective envelope bottom, we use the values of the Airy functions
and their derivatives at $s=0$,
$${\cal A}i={{\cal B}i\over\sqrt3}, \quad {\rd{\cal A}i\over \rd
s}={-1\over\sqrt3}{\rd{\cal B}i\over \rd s}, \quad{1\over{\cal
B}i}{\rd{\cal B}i\over \rd
s}=3^{1/3}{\Gamma(2/3)\over\Gamma(1/3)}=0.729,$$ and $A(r_b)=0$.

At $r=r_b$, with the use of Eqs. 10, 17, 22, and 23, we finally
obtain
\begin{equation}
 \left(\matrix{y_1\cr
y_2\cr y_3\cr y_4}\right)= \tilde C_s\left(\matrix{1\cr\ 1\cr -1+B\cr
2-U+B}\right) +\tilde C_r\left(\matrix{2{\cal Z}_1\cr {\cal Z}_2\cr
-UB{\cal Z}_1\cr -UB[{\cal Z}_2+(U-4){\cal Z}_1]}\right),
\end{equation}
where
$$\tilde C_s={C_s\over r_b},\qquad \tilde C_r={C_r\over(r^3\sqrt{\rho})_b},
$$
$${\cal Z}_1=1-{{\rm i}\over\sqrt3},\quad\mbox{and}\qquad
{\cal Z}_2=0.729\beta\left(1+{{\rm
i}\over\sqrt3}\right)-{V_g\over2}{\cal Z}_1.$$ The same form is
valid for the inner boundary condition applied within the radiative
core but, as follows from Eq. 19, with  ${\cal Z}_1=1$, and
$${\cal Z}_2=\left({\rm i}k_r+{A_-V_g\over2}\right){\cal Z}_1.$$

Two inner boundary conditions in the form of linear algebraic
relations are obtained by eliminating constants $\tilde C_s$ and
$\tilde C_r$. The relations are complex and so are the resulting
eigenvalues and eigenfunctions even if, as we will do in this paper,
all nonadiabatic effects in the envelope are ignored.

Derivation of the outer boundary conditions for oscillations in realistic stellar models cannot be done in
such a rigorous way as in the case of polytropic models considered by Takata (2005). In realistic models,
 $V_g(R)<\infty$, so there is no singularity, and $U>0$. In main sequence stars similar to the Sun,
$V_g\sim A\sim10^3$ and $U\sim10^{-6}$. In stars near the tip of the
RGB, $V_g\sim A\sim10^2$ and $U\sim10^{-3}$. The standard procedure
is to set the boundary condition in the atmosphere near maximum of
$V_g$ and ignore $U$. Then in an equation like  Eq. 7, the
coefficients are approximately constant and for the wave reflected
downward we have
\begin{equation}
By_2=\left[1-{\Gamma_1\over2}\left(1-\sqrt{1-\left({\omega\over\omega_{\rm ac}}\right)^2}\right)\right]y_1,
\end{equation}
where $\omega_{\rm ac} \equiv g\Gamma_1/2c$ is the acoustic
frequency. The vacuum solution of the Poisson equation outside the
star implies  $y_4=-2y_3$ at $r=R$. This is the second outer
boundary condition adopted in this paper. For dipolar modes, an
additional condition follows from Eq. 4, which combined with the
previous two yields $y_3=U(y_1-y_2)/3$. Thus, we have three
conditions and, with the standard normalization $y_1=1$, we may
carry a single integration from the surface downward. Application of
this standard procedure to luminous red giants may seem
problematical and therefore the sensitivity of results to variation
in the outer boundary condition must be checked. Only one of the
inner boundary conditions is needed in a search for eigenfrequency.
The second condition may be used as a test of numerical accuracy.

\section{The acoustic propagation zone}

It is well-known that frequencies of low order radial modes in
evolved stars may be accurately calculated from their envelope
models alone.  We will see that in the case of luminous red giants,
this property extends up to the highest orders. The bottom of the
envelope, which must cover the whole acoustic propagation zone, may
still be within the convective region. Thus, what determines the
depth of the propagation zone is not the ${\cal N}(r)=\omega$
condition, where ${\cal N}$ is the Brunt-V\"{a}is\"{a}l\"{a}
frequency. A general condition limiting the propagation zone from
its bottom does not exist\footnote{The generalized acoustic cut-off
frequency, $\omega_c$, as defined e.g. in (Aerts et al. 2010) is not
a good choice in the present application because $\omega_c^2<0$ in
the lower part of the envelope.}.  The condition $\omega<\omega_{\rm
ac}$ for wave reflection is derived for an isothermal plane-parallel
atmosphere built of perfect gas. Although in this case the condition
is numerically close to $\omega<{\cal N}$, its origin is different
(to see this consider, e.g., the case of gas with $\Gamma_1=1$ when
${\cal N}=0$). In fact, the condition follows from general rule
prohibiting wave propagation if the background structure changes
significantly over wavelength. The equality $\omega=\omega_{\rm ac}$
corresponds to equality of the radial wave number acoustic waves to
$0.5/H_p$, where $H_p=p/g\rho$ is the pressure distance scale. It
does not seem unreasonable to use $\omega<\omega_{\rm ac}(r)$ as a
crude local no-propagation condition, as long as $H_p$ is the
shortest of all local distance scales.

For nonradial modes the acoustic propagation zone may be limited
from the bottom by the Lamb frequency, ${\cal L}_\ell=\sqrt{\ell(\ell+1)}/r$ exceeding $\omega$.
This critical frequency was determined upon assuming the Cowling approximation.
Avoiding this approximation, Takata (2006b) found a simple modification of the Lamb frequency for dipolar modes,
which in our notation may be written as
$${\cal L}_{1m}={\cal L}_1\left(1-{U\over3}\right).$$
In our application, this modification is not very significant. Near
the bottom of convective envelope the correction is at most one
percent. The local maximum of $U/3$ (up to 0.5) is reached in the
middle of the envelope. We will not use modified ${\cal N}$, where the
same factor appears in the denominator, because it has no effect on
lower boundary of the acoustic propagation zone.

During evolution along the RGB frequencies of stochastically excited
oscillations decrease. The critical frequencies $\omega_{\rm ac}(r)$
and ${\cal L}_{1m}(r)$ at specified fractional radius decrease too,
but not as fast. Therefore the bottom of the acoustic propagation
zone moves upward. In the same time, the bottom of the convective
envelope moves downward. This leads to expansion of the evanescent
zone between the gravity and acoustic propagation zones and to
efficient mode trapping in the convective envelope.

As an example. we consider a sequence of red giant envelope models
with mass, $M$, luminosity, $L$, and effective temperature, $T_{\rm
eff}$, taken from BaTSI (Pietrinferni et al. 2006) evolutionary
tracks for the initial mass $M_0=2M_\odot$ and heavy element abundance
parameter $Z_0=0.02$. The opacity and equation of state data come
from the same source. The adopted mixing length parameter
$\alpha=1.85$ corresponds to 1.74 in BaTSI models. Parameters of
these models are listed in Table 1. The models may be regarded
realistic down to the bottom of the convective envelope. Listed in
the table values of  $\nu_{\rm max}$, corresponding to the expected
maximum of acoustic power for stochastic oscillations, were evaluated with the Kjeldsen
\& Bedding (1995) expression
\begin{equation}
\nu_{\rm max}={M\over L}\left({T\over T_\odot}
\right)^{3.5}\times3050\mu Hz.
\end{equation}
The values of $\Delta\nu$, which denote the distances
between consecutive  radial modes (the large separations), were calculated
around  $\nu_{\rm max}$ for reconstructed envelope models with
Warsaw codes (see e.g. Dziembowski \& Soszy\'nski 2010).
Models 1 to 5 represent stars ascending the
RGB.  Model 6 in Table 1 represents a helium burning (red clump) object
\begin{table}\caption{Selected models from the BaSTI track at $M_0=2M_\odot$ and
$Z=0.02$.}
 \label{table:1} \center
\begin{tabular}{c c c c l l l }     
\hline &$M$ & $\log L$ & $\log T_{\rm eff}$ & $r_b/R$ & $\nu_{\rm max}$ & $\Delta\nu$ \\
&$(M_\odot)$&$(L_\odot)$&(K)&&$(\mu$Hz)&$(\mu$Hz)\\
\hline
   1 & 1.998 & 1.500 & 3.692 & 0.0746 & 110.8 & 8.851\\
   2 & 1.997 & 2.000 & 3.663 & 0.0454 & 27.47 & 3.071\\
   3 & 1.994 & 2.501 & 3.628 & 0.0287 &  6.51 & 1.050\\
   4 & 1.985 & 3.001 & 3.586 & 0.0104 &  1.46 & 0.329\\
   5 & 1.967 & 3.355 & 3.552 & 0.0061 &  0.49 & 0.137\\
   6 & 1.967 & 1.771 & 3.687 & 0.2410 & 55.90 & 5.752\\
   \hline
\end{tabular}
\end{table}

We begin with the extreme case, that is, with Model 5, which is
located at the tip of red giant branch. The three critical
frequencies in this model are shown in the top panel of Fig. 1. The
acoustic cavity for radial mode is determined solely by the run of
$\omega_{\rm ac}$. For dipolar modes the bottom of the acoustic
propagation zone is determined by
$$\omega=\max[\omega_{\rm ac}(r),{\cal L}_{1m}(r)]$$
condition and the top of the gravity propagation zone by
$\omega={\cal N}$. In the middle and bottom panels, we show
fractional inertia
\begin{equation}
I(r)=\int_{r_b}^r(|y_1|^2+2|y_2|^2)\rho r^4\rd r
\end{equation}
for radial and dipolar modes, respectively. For these plots,
normalization $I\equiv I(R)=1$ was adopted. With the standard
normalization $y_1(R)=1$, the value of $I\equiv I(R)$ is a measure
of mode trapping. Let us stress that $I$ is not the total mode
inertia but only its part contributed by the convective envelope.
The plots may suggest that for any of the modes shown in Fig. 1,
there is no contribution to the total inertia (energy) from the
interior below $r\approx0.2$. In fact, for the dipolar modes the
inertia, $I_c$, arising from the slow component, which keeps the
center of mass of the whole star at rest, is substantial. This
contribution to the total inertia may be calculated without any
knowledge of the core structure. Using in Eq. 27 the expressions for
$y_1$ and $y_2$ from Eq. 10,  we get
\begin{equation}
I_c\equiv\int_0^{r_b}(|y_{s,1}|^2+2|y_{s,2}|^2\rho r^4\rd r=3\left({\rho r^5\over U}\right)_b|\tilde C_s|^2.
\end{equation}
The $I_c/I$ ratio ranges up to 1/3 for the $p_{0,1}$ mode and decreases with the mode order.

According to the terminology adopted in works on stellar pulsation,
the radial p$_1$ is called fundamental (no nodes between the surface
and the lower boundary and one zone of steep rise in $I$). Its
frequency is well below the minimum value of $\omega_{\rm ac}(r)$.
For the overtones, most of the contribution to $I$ arises in the
zone  $\omega\lesssim\omega_{\rm ac}(r)$ condition or close to it.
Comparison of middle and bottom panels reveals first of all a
similarity of the energy distribution in dipolar and radial modes,
particularly in case of first (fundamental) modes whose frequencies
are close. More differences are seen for the overtones.

Let us note in the top panel that the critical frequencies
$\omega_{\rm ac}(r)$ and ${\cal L}_{1m}(r)$ are close to each other
in the frequency range of the dipolar overtones depicted in the
panel. The role of the Lamb frequency is reflected in the shape of
$I(r)$ for all dipolar overtones. Beginning with p$_3$, this
critical frequency determines the bottom of the acoustic propagation
zone.

 \begin{figure}
   \centering
   \resizebox{\hsize}{!}
   {\includegraphics{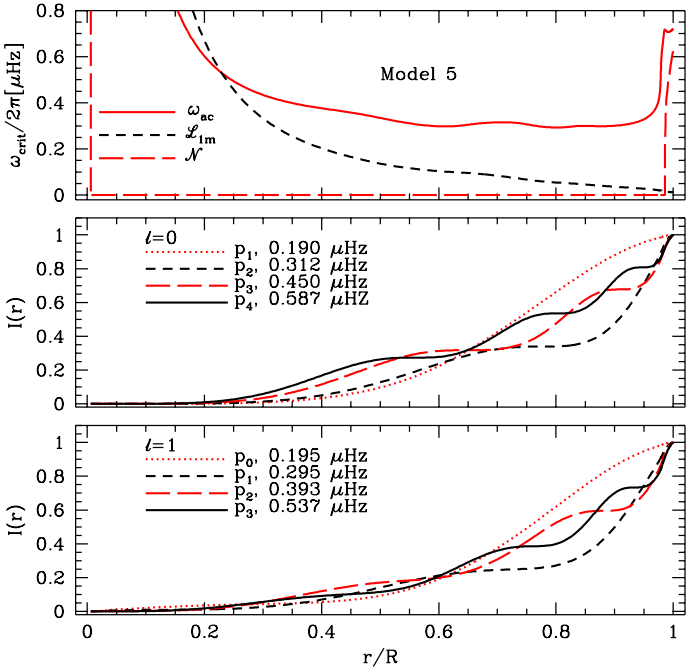}}
   \caption{Critical frequencies (see text) in Model 5 (see Table 1).
   ({\it top}) Relative contributions to mode inertia integrated upward from the
   base convective envelope for all four modes with frequencies below $\omega_{\rm
   ac}(R)$. ({\it middle}) The same for the first four dipolar modes trapped in
   the envelope. ({\it bottom}) There is one more dipolar overtone,
   whose frequency $\nu=0.677 \mu{\rm Hz}$ is below $\omega_{\rm ac}(R)$.}
   \label{Fig1}%
\end{figure}

Where between consecutive radial modes the dipolar mode appears depends
on the critical frequencies determining the lower boundary of the
acoustic propagation zone. If it is $\omega_{\rm ac}$, which is
$\ell$-independent, the dipolar and radial modes have close
frequencies. This is the case for the low order modes in Model 5, as
we may see in numbers quoted in Fig. 1. With increasing oscillation
frequency, the role of the  Lamb frequency increases and the dipolar
mode moves away but still is rather far from the mid point between consecutive
radial modes.  It comes closer to this point in less luminous stars
where higher order modes occur near
$\nu_{\rm max}$.

Fig. 2 shows in separate panels fractional moment of inertia for
modes with frequencies around $\nu_{\rm max}$ in Models 1 to 4. The
places where the critical frequencies are equal to $\nu_{\rm max}$
are marked in each panel. The bottom of the acoustic cavity is
clearly within the convective envelope except for Model 1, where the
equality  $\omega_{\rm ac}=2\pi\nu_{\rm max}$  takes place in the
radiative core. For the second critical frequency, the equality
occurs in the envelope but the evanescent zone is too narrow for an
efficient mode trapping of the dipolar mode within the convective
envelope. We may observe in this figure how the width of the
evanescent zone increases with stellar luminosity leading to
trapping $\ell=0$ and 1 modes in the upper  part of the envelope.
This is seen in the upward shifts of the oscillatory pattern in
$I(r)$.

 \begin{figure}
 \centering
   \resizebox{\hsize}{!}
   {\includegraphics{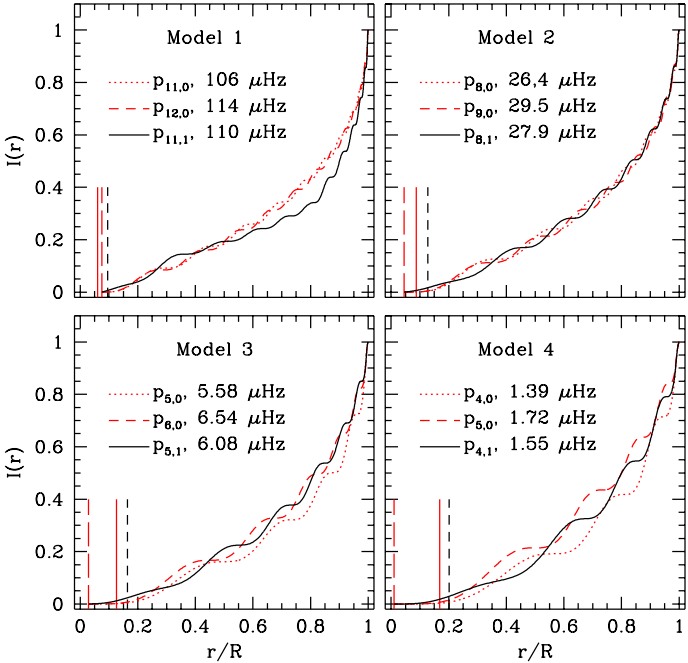}}
   \caption{Relative contributions to mode inertia integrated upward from the
   base convective envelope for three modes (two radial and one dipolar)
   with frequencies close to $\nu_{\rm max}$ in the four indicated models
   whose parameters are given in Table 1. The short lines parallel to the ordinate
   axis mark the bottom of the convective envelope (long-dashed style),
   the places where $2\pi\nu_{\rm max}=\omega_{\rm ac}$, and ${\cal L}_{1m}$ (drawn
   in the solid,  and short-dashed style, respectively). The styles correspond to those used to show
   the respective frequencies in Fig. 1.}
   \label{Fig2}%
\end{figure}

 \begin{figure}
   \centering
   \resizebox{\hsize}{!}
   {\includegraphics{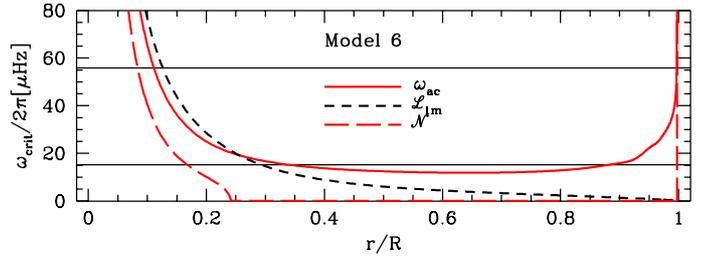}}
   \caption{Critical frequencies in Model 6. The two horizontal lines correspond to dipolar mode
   frequencies. The lower one to p$_{1,1}$ and the upper to p$_{9,1}$, whose frequency is near $\nu_{\rm max}$.}
   \label{Fig3}%
\end{figure}

Fig. 3 shows critical frequencies in Model 6 representing an object
in the phase of He burning in the convective core. The convective
envelope is relatively shallow and the ${\cal L}_{1m}=2\pi\nu_{\rm
max}$ equality takes place in the radiative part of the interior. In
Sect. 3, we described the version of the running wave boundary
condition, which may be applied in the gravity wave propagation
zone. However, it is rather unlikely that the assumption that the
reflected wave has negligible amplitude might be justified in this
case.

\section{Mode trapping}
Mode trapping in the acoustic cavity is a result of
wave reflection within the evanescent zone. The efficiency of this
process depends on the width of this zone and on mode degree.
The eigenfunction in the evanescent zone is a superposition of monotonically
increasing and decreasing components.
At higher degrees, the dependence of the displacement on the distance to the  center
in the two components is approximately given by $r^{\ell-1}$ and $r^{-(\ell+2)}$.
The large contrast between these behaviors causes one of the component to dominate and
it is the increasing component for modes trapped in the envelope.
A narrow evanescent zone suffices for an efficient trapping. We may speak about hard
reflection.
This effect is behind the occurrence of unstable modes of
moderate $\ell$ in models of Cepheids and RR Lyrae stars.
At $\ell=0$ and 1 there are also  monotonically increasing and decreasing
components but not as fast as at higher degrees.
The reflection is softer and a wider evanescent zone is needed for an efficient trapping.

For radial modes, efficient trapping manifests itself in a weak
dependence of frequencies on the exact form of the inner boundary
condition set at the bottom of the convective envelope. This is so
because then the contribution from the decreasing component becomes
very small once the acoustic cavity is reached and thus have a
negligible influence on eigenfrequencies. The effect may be
quantified by considering the generic form of the boundary condition
for homogenous second order equations
\begin{equation}
\cos(\pi\phi)r{{\rm d}y_1\over{\rm d}r}+\sin(\pi\phi)y_1=0
\end{equation}
and letting $\phi$ vary between 0 and 1. The numbers quoted in Figs.
1 and 2 were obtained assuming $\phi=0$. In Fig. 4 the effect is
illustrated for  Models 1 and 2. In the former model, except of a
narrow range, the value of $\phi$ has almost no effect on
frequencies. In this range, the  boundary condition allows a large
decreasing component which is certainly impossible for any realistic
model of the whole star. With the $y_1(r_b)=0$ condition we are safe
away from this bad range. In Model 1 the effect of varying $\phi$ is
considerably larger which is consistent with what we see in Fig. 2.
In this model, the propagation zone for the two radial modes extends
to the core.

 \begin{figure}
 \centering
   \resizebox{\hsize}{!}
   {\includegraphics{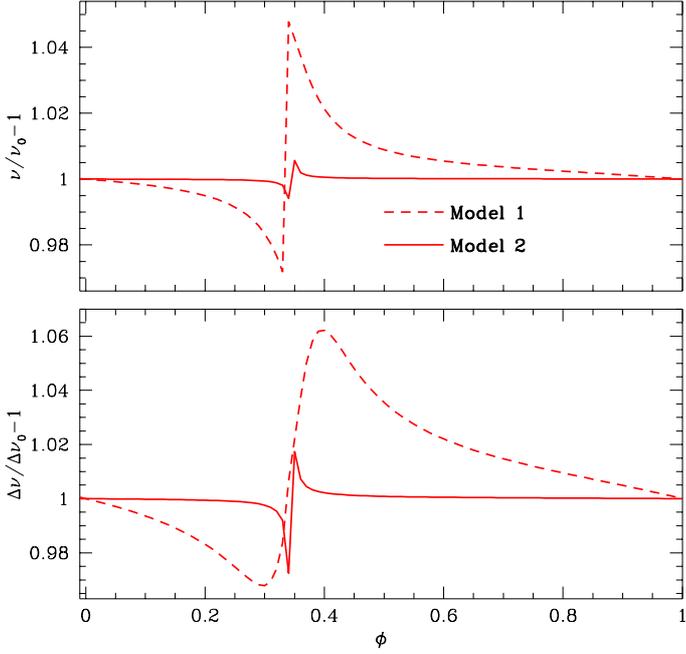}}
   \caption{Dependence of radial mode frequencies, $\nu$, on the parameter $\phi$ adopted in Eq. 29. The pairs of frequencies
   around $\nu_{\rm max}$ in Models 1 and 2 are considered. Plotted are the shifts relative to the values at $\phi=0$.
   The top panel shows the shift in the lower frequency. The bottom panel shows the shift in the frequency difference.}
   \label{Fig4}
\end{figure}

Changes in dipolar mode trapping may be observed by relaxing the
inner boundary condition and treating $I$, defined in Eq. 27, as a
continuous function of frequency. The top panel of Fig. 5 shows the
values of $I$ in the frequency range centered at $\nu_{\rm max}$ for
Models 1 and 2. The inertias of radial modes in the same frequency
range are shown for a comparison. It is important to notice the
difference between the two models. The minima for Model 2 are far
deeper than in Model 1. The difference in trapping is better visible
in the lower panel showing values of $|y_1+y_3|$ at the lower
boundary. In this combination of the eigenfunctions, the
contribution of the slow component in $\xi_r$ is nearly eliminated,
as we may see in Eq. 24. In this case the minimum value is zero. The
dependence is so steep in the case of luminous red giants that
unfitted envelope models suffice to determine frequencies of
expected dipolar mode frequencies at specified stellar parameters.

The transition to efficient trapping occurs at $\log L\approx2$.
Luminosity of our Model 1 places it between Models A and B
of Dupret et al. (2009) while that of Model 2 between their Models B
and C. In the latter model ($\log L=2.1$) they find that damping in radiative
interior is high enough to reduce amplitudes of all dipolar modes
 below the detection level, except the ones corresponding to $I$ minima.
The value of the critical luminosity refers to models
with $M_0=2$ and it is lower at lower mass.

 \begin{figure}
 \centering
   \resizebox{\hsize}{!}
   {\includegraphics{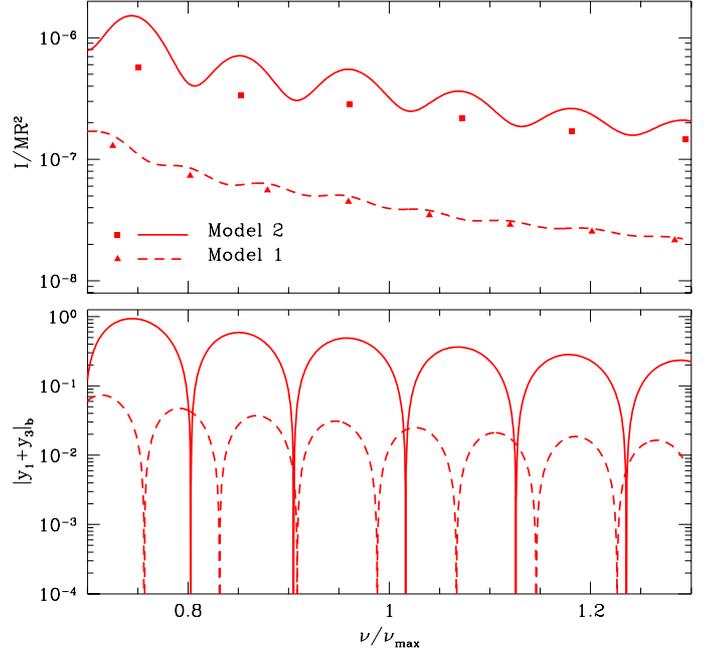}}
   \caption{Inertias of dipolar and radial modes in the frequency range centered at
    $\nu_{\rm max}$ in Models 1 and 2 are shown in the top panel.
    For dipolar modes the inner boundary condition was relaxed so that inertia is a continuous
    function of frequency.
    The standard outer boundary condition and normalization were adopted in all the cases.
     The quantity plotted in the bottom panel is a measure of the gravity wave
    amplitude at the top of the propagation zone.}
   \label{Fig5}%
\end{figure}

\section{Frequencies of the trapped modes}

The minima of $|y_1+y_3|_b$ shown in Fig. 5 occur at frequencies
close to real part of the eigenfrequencies obtained with the code
described in Sect. 3. The location of the minima and the dipolar
mode frequencies close to the mid points between consecutive radial
mode frequencies is expected in the $B\rightarrow\infty$ limit. Note
that in this limit $S\rightarrow0$ and Eq. 7 takes again the form
corresponding to the Cowling approximation, which is the basis of
the usual asymptotic expressions for dipolar mode frequencies.
However, asymptotic expressions  yield a good approximation for
eigenfrequencies only if at the place where ${\cal L}_1(r)=2\pi\nu$,
we have $|S|\ll1$ and this is not true in most the cases considered
in this paper. Beginning with  $\ell=2$, the Cowling approximation
is applicable and the asymptotic expressions may be derived in the
usual way.

 \begin{figure}
 \centering
   \resizebox{\hsize}{!} {\includegraphics{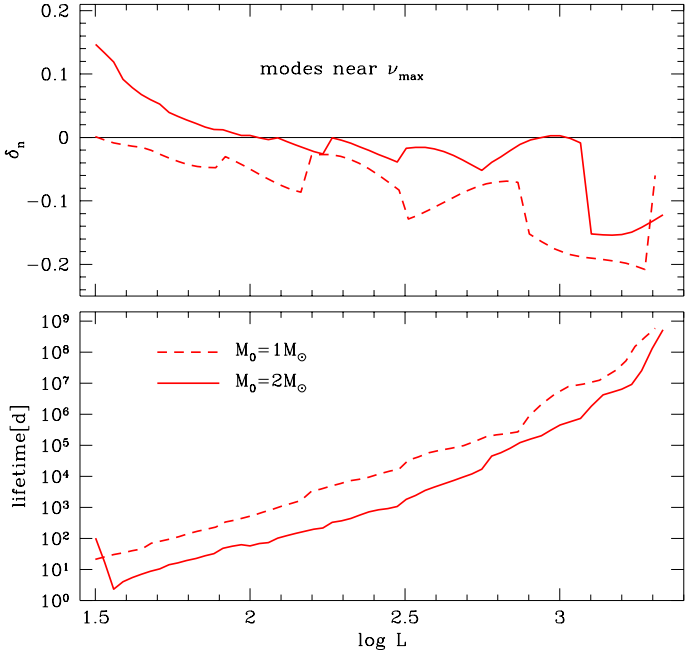}}
   \caption{
   The normalized small separation defined in Eq. 30 (top) and
   the lifetimes limited only by wave energy loss (bottom) for the
   dipolar mode with frequencies closest to $\nu_{\rm max}$ in
   red giant envelope of indicated luminosity and initial mass. }
   \label{Fig6}
\end{figure}

 \begin{figure}
 \centering
   \resizebox{\hsize}{!}
   {\includegraphics{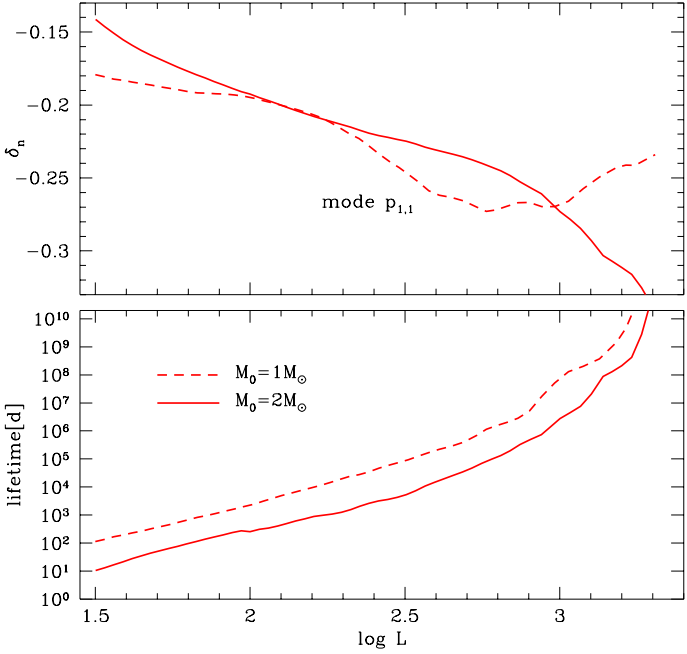}}
   \caption{The same as in Fig. 6 but for p$_{1,1}$ mode}
   \label{Fig7}%
\end{figure}

The real parts of eigenvalues obtained in the way described in Sect.
3 differ very little from eigenfrequencies obtained with the simple
real inner boundary condition $y_1(r_b)=0$.  Only in Model 1 the
difference exceeds one percent.

The small separations based on dipolar and radial mode frequencies are one of standard
characteristics of stochastic oscillations. Here,
we use relative small separations defined as
\begin{equation}
\delta_n\equiv{\delta\nu_{0,1}\over\Delta\nu}={0.5(\nu_{n+1,0}+\nu_{n,0})-\nu_{n,1}\over\nu_{n+1,0}-\nu_{n,0}},
\end{equation}
which  measure the departure of dipolar mode frequencies from the
mid position between consecutive radial modes. In the top panels of
Figs. 6 and 7 we show values of $\delta_n$ for modes at
$\nu\approx\nu_{\rm max}$ and  $n=1$, respectively. The values were
calculated for envelope models with the surface parameters taken
from BaSTI tracks for metallicity parameter $Z=0.02$ and initial
masses $M_0=1$ and $2M_\odot$. In Table 2  we list selected parameters for
selected models at $M_0=1M_\odot$. Similar data for models at $M_0=2M_\odot$ and
nearly the same $L$ were given earlier in Table 1.

\begin{table}\caption{Selected models from the BaSTI track at $M_0=1M_\odot$ and
$Z=0.02$.}
 \label{table:2} \center
\begin{tabular}{c c c c l l l }     
\hline &$M$ & $\log L$ & $\log T_{\rm eff}$ & $r_b/R$ & $\nu_{\rm max}$ & $\Delta\nu$ \\
&$(M_\odot)$&$(L_\odot)$&(K)&&$(\mu$Hz)&$(\mu$Hz)\\
\hline
   1 & 0.997 & 1.500 & 3.658 & 0.0628 & 41.53 & 4.869\\
   2 & 0.994 & 2.000 & 3.628 & 0.0410 & 10.31 & 1.717\\
   3 & 0.986 & 2.501 & 3.590 & 0.0213 &  2.38 & 0.586\\
   4 & 0.964 & 3.005 & 3.542 & 0.0125 &  0.49 & 0.172\\
   5 & 0.933 & 3.275 & 3.511 & 0.0132 &  0.20 & 0.092\\
   \hline
\end{tabular}
\end{table}

The jump-like changes, visible in the first figure, result from
gradual switches of mode orders,  from $n=8$ to $3$ and from $n=12$
to 4 for $M=1$ and $2M_\odot$, respectively. The changes at constant $n$, as
we may see in Fig. 6, are smooth but  not necessarily monotonic. At
$n=1$, the dipolar mode occurs always significantly closer to the
higher frequency radial mode. Clearly, negative values of $\delta_n$
prevail also for modes near $\nu_{\rm max}$. Only in the low
luminosity models at $M_0=2M_\odot$ the sign is opposite. This agrees with
results of  Montalban et al. (2010). Negative values of $\delta_n$
were also found in the fit to the CoRoT data (Mosser et al.
2010). Significant departure from the mid position means that
$\delta_n$ is an independent seismic observable, which may perhaps
be measured with future, more precise data. As for today, only mixed
dipolar modes give us new seismic information about red giants
beyond $\Delta\nu$ and $\nu_{\rm max}$. We will return to this
matter in Section 8.

\section{Effect of wave losses on mode lifetime}

The imaginary part of the eigenfrequency calculated with our code yields the damping rate, $\gamma=-\omega_I$,
 which takes into account only the effect of the wave losses at the bottom of the convective envelope.
Of course, this is not the only important source of damping but the
other resides predominantly in the outermost layers, where
$\omega_{\rm ac}\gg{\cal L}_\ell$ and the eigenfunctions depend
mainly on mode frequency and not on its degree. Therefore, the
difference between dipolar and radial mode damping rates may be
assessed without explicit calculation of nonadiabatic effects in
these layers.

Proceeding a similar way as in derivation of the variational
expression for real eigenvalues, we begin with multiplying Eq. 1 by
$\vxi^*$ and integrating over the envelope. In further
transformation, we make use of the Poisson and continuity equations
and finally arrive at
 \bee
\omega^2I&=&-\int_{\Sigma_b}\left[\xi_r^*(p'+\rho\Phi')+ {\Phi'\over4\pi
G}{\partial \Phi'*\over\partial r}\right]\rd\Sigma
+\nonumber\\&&\int_V\left({|p'|^2\over p\Gamma_1}+
Ag\rho{|\xi_r|^2\over r}-{|\vnab\Phi'|^2\over4\pi G}\right)\rd^3\vx - E,
\ene
where $E$ is the {\it ad hoc} added $\ell$-independent term describing the
near surface effects that may contribute to damping (or driving).

Assuming $|\omega_I|\ll\omega_R$, we get from the imaginary part of
Eq. 31

$$
\omega_I=-{W+E_I\over2\omega_R I},$$
 where
$$W=\int_{\Sigma_b}\Im\left[\xi_r^*(p'+\rho\Phi')+
{\Phi'\over4\pi G}{\partial \Phi'*\over\partial r}\right]\rd\Sigma,
$$
which in terms of the radial eigenfunctions defined in Eqs. 2 and 3
becomes
$$W=(\rho r^5)_b\left[\Im(\omega^2y_1^*y_2)+{g\over rU}\Im(y_3y_4^*)\right]_b.$$
With the use of the asymptotic solution given in Eq. 24,  a more
revealing expression for the damping rate caused by the wave losses
may be obtained. Neglecting terms of the orders of $B$ and
$\omega_I/\omega_R$ in comparison with 1, we get
$$W=(\rho r^5)_b\left[2\omega_R^2\Im({\cal Z}_1^*{\cal Z}_2)|\tilde C_r|^2
+{6\omega_I\omega_R\over U}|\tilde C_s|^2\right].$$ The leading
terms in the coefficient at the $\tilde C_s\tilde C_r$ product
cancel out. From Eqs. 14 and 16, we have
$$\Im({\cal Z}_1^*{\cal Z}_2)=0.842\beta.$$
Thus,  the total damping rate, $\gamma=-\omega_I$, is given by
$$
\gamma=\gamma_W+\gamma_E=-\omega_I={1\over I+I_c}\left(
0.842\omega_R|\tilde C_r|^2\beta(\rho
r^5)_b+{E_I\over2\omega_R}\right).
$$
The expression for $I_c$, which is the part of mode inertia arising
in the core, is given in Eq. 28. The rapid component contributes
only to the energy loss because the reflected wave was neglected.

The near surface effects have some influence on the eigenfunctions
at the bottom of the envelope and, hence, on the value of $W$. This
influence was assessed by adding to the coefficent at $y_1$ in the
r.h.s. of Eq. 25 an imaginary number leading to the values of
$\gamma_E$ of the order of those found by Dupret et al. (2009). The
changes in values of $\gamma_w$ were found small enough to regard
the two dominant sources of damping as additive.

Damping due to wave losses sharply decreases with increasing red
giant luminosity. This is shown in the bottom panels of Figs. 6
and 7 for the dipolar modes with frequencies closest to $\nu_{\rm
max}$ and for p$_{1,1}$ modes, respectively. The plots depict, for a
number of envelope models, the values of $\tau_W=\gamma_W^{-1}$,
that is, lifetimes limited only by the wave emission. There is some
uncertainty in evaluated lifetimes following from the uncertainty in
the derivative  $\rd A/\rd r$, at the top of radiative interior.
This derivative enters the inner boundary condition (see Eq. 24)
through parameter $\beta$ defined in Eq. 21. Our standard approach
is to evaluate $\omega$ and $\tau_W$ assuming constant element
abundance but the He abundance may rise immediately beneath the
bottom of the convective envelope. To check consequences of this
uncertainty, calculations were performed with the derivative
increased by factor 10. As expected, the effect
 on frequencies was negligible. The lifetimes became by factor about two longer. This is not much for such a drastic modification
of the $A$ derivative and, especially, when the effect is compared with the four orders of magnitude increase of $\tau_W$ between $\log L=2$ and 3.
The general pattern of the $\tau_W(L)$ dependence, which we see in these two figures, are
similar. The lifetimes are longer for the p$_{1,1}$ modes because at lower frequency the evanescent zone is wider, hence
trapping is more efficient, and the wave amplitudes are lower. Decreasing mass at a specified luminosity acts in the same
direction.

Let us now focus on modes with frequencies $\nu_{\rm max}$ which are
of interest for comparison with observations. The lifetime due to
wave losses increases rapidly with $L$ and it scales roughly as
$M/L$. The average exponent, $w$, in the $\tau_W\sim L^w$ relation
is close to 4. It is interesting to compare the role of wave losses
and the near surface damping in limiting mode lifetimes. For the
latter effect we use values of $\tau_E$ calculated for radial modes
of by Dupret et al.(2009). They calculated also lifetimes for
nonradial modes but these depend on mode properties in the core.
Therefore, it seems safer to rely on results radial modes at similar
frequencies. For the trapped dipolar modes, which we want to
consider, $I_c/I$ is small and $I$ is close to mean value of the
nearest radial modes. So that, we may use mean of the lifetimes of
these modes as a crude assessment of $\tau_E$ for trapped dipolar
modes.

In models with $M_0=2M_\odot$ at $\log L=1.8$ and 2.1, the values of
$\nu_{\rm max}$ are 49 and 21$\mu$Hz , respectively. From our Fig.
6, we find  the values 20 and 110 d for $\tau_W$. The corresponding
values of $\tau_E$ inferred from Figs. 7 and 8 of Dupret at al. are
20 and 30 d. Thus, we may conclude that at $M_0=2M_\odot$ and $\log
L\gtrsim2$ the wave losses become a secondary source of damping and
become negligible at somewhat higher luminosity. At lower initial
mass the same happens at accordingly lower luminosity. At $M_0=2M_\odot$
and $\log L=2.1$ the total relative damping rate,
-${\omega_I/\omega_R}=(\nu_{max}\tau)^{-1}=0.024$.

 With this estimate, we may return to the matter of validity of the running wave approximation,
 discussed in Sect. 2. In order to use condition
given in Eq. 18, we still need the values of $\Psi_R$ and $\int D\rd
\Psi_R$ for which  complete stellar models. For the present estimate
we will use numbers for the $M=2M_\odot$ models calculated for Dziembowski
et al. (2001) paper which should be adequate for this aim. The
values of $\Psi_R$ and $\int D\rd \Psi_R$ at $L=2$  are
$1.9\times10^3$  and 0.49, respectively, while at $L=2.5$ the
corresponding values are $1.0\times10^4$ and 370 (in this case the
required exact expression for $k_r$ was used). The value of $\int
D\rd \Psi_R$ grows much more rapidly with $L$ than $\Psi_R$ and it
is clear that
 at the value slightly
exceeding 2 the running wave approximation is justified even if only the linear radiative damping is considered.

\section{Discussion}

Solar-like oscillations in low luminosity red giants are interpreted as global mixed modes with
both the gravity and acoustic cavities contributing, in various proportions, to the total inertia.
At higher luminosity, a more adequate picture is that of acoustic modes of the convective
envelope loosing  energy via gravity wave emission from its bottom.
With growing luminosity this loss becomes insignificant.

Transition between these two pictures is continuous but very sharp.
From Dupret et al. (2009) and the arguments presented in this paper
it follows that for stars of the initial mass $M_0=2M_\odot$ the transition
occurs near luminosity $L\gtrsim100$, which corresponds (see Table
1) to $\nu_{\rm max}=27.5$ and $\Delta\nu=3.1\mu$. At lower mass,
the transition takes place at lower luminosity but the  values of
$\Delta\nu$ at the transition should be similar. In Fig. 3 of Mosser
et al. (2010) most of data points lie above $\Delta\nu=3\mu$Hz but
there are also lots of data points below $\Delta\nu=2\mu$Hz  which
are most likely associated with luminous objects, where only
acoustic envelope modes are expected. The most striking feature
shown in this figure is the universal pattern of the relative
positions of the dipolar and radial modes across the whole range of
$\Delta\nu$ values from 1 to $8 \mu{\rm Hz}$. The dipolar modes are
located between radial modes somewhat shifted off the mid point
toward high frequency. This is consistent with what is seen in Fig. 5 for
the envelope modes. We may also see that some spread of
the position of such modes is expected. Similar situation is
expected for the mixed dipolar modes corresponding to minima of the
inertia (Mont\'alban et al. 2010).

A convincing evidence for the mixed mode excitation comes from
detection of few dipolar modes near the mid points between
consecutive radial modes (Beck et al. 2011a, Bedding, et al. 2011,
Mosser et al. 2011). In the first two papers, three objects from the
Kepler catalog are discussed. The values of $\Delta\nu$ are in the
range of 7.9 to $10 \mu{\rm Hz}$ where stochastic excitation
of mixed modes is indeed expected. However, Mosser et al. (2011) in
their analysis of the CoRoT data find a large number of such objects
in a wide range of $\Delta\nu$ extending  down to 2 $\mu$Hz, that is
well below the range expected range of mixed modes. They also report
an evidence for mixed $\ell=2$ mode excitation and note this is in
conflict with the  Dupret el. (2009) prediction that such modes
should be eliminated by strong damping in the core. A possible
solution of these two problems could be the gravity  wave reflection
caused by a steep composition gradient  above the hydrogen burning
shell, where most of the damping takes place. The large values of
$\Delta T_{\rm obs}$  determined by Mosser et al. for the dipolar
mixed modes at low $\Delta\nu$ are consistent with such an
interpretation.

Acoustic envelope modes are totaly ignorant of what happens below
the convective envelope. Their frequencies yield independent but
similar information to radial modes because in both cases,
frequencies are determined by sound speed distribution within the
envelope. Only mixed modes  probe the radiative interiors of red
giants. Data on multiple peaks has been already used to distinguish
between helium burning and other red giants (Bedding et al. 2011,
Mosser et al. 2011). There is a prospect for going beyond that. In
particular, we may get interesting constraints on element
distribution from frequencies and on dissipative processes in the
core from amplitudes and lifetimes. A new exciting result based
on mixed dipolar modes is the evidence for fast rotation in red giant
cores (Beck et al. 2011b).

We do not yet have data satellite on oscillation in most luminous
red giants ($\log L\gtrsim2.5$, $\Delta\nu\lesssim1$). Rich data on
such objects located in the Magellanic Clouds are available from
OGLE project (Soszy\'nski et al. 2009). The advantage of these data
is that luminosity of individual objects is known to a good
accuracy. The peaks occur at frequencies below $1\mu{\rm Hz}$, which
is consistent with prediction based on Eq. 26. The frequencies are
in the range of lowest order p-modes. In a number of cases two
clearly dominant peaks are seen. However, the frequency separations
are by some 20\% lower than expected large separation. This could
not be explained in terms of two $\ell=0$ modes, nor in terms of the
$\ell=0, 1$ pair (Dziembowski \& Soszy\'nski 2010). In the light of
the discussion in the present paper the second option would appear
more plausible but, in fact, it leads to a greater discrepancy.
Satellite data on such objects could provide us a key to the
solution of this intriguing problem.

\begin{acknowledgements}
I am grateful to Masao Takata for reading a preliminary version of this paper and
suggesting a number of corrections.
This work was partially supported by the Polish MNiSW grant number N
N203 379636.
\end{acknowledgements}
{}

\begin{thebibliography}{}
\bibitem[]{} Abramovitz, M. \& Stegun, I.~A. ~(eds.) 1972,
Handbook of Mathematcal Functions (John Wiley and Sons, New York)
\bibitem[]{} Aerts, C., Christensen-Dalsgaard, J., \& Kurtz, D.~W 2010
 Asteroseismology (Springer, Heidelberg)
\bibitem[]{} Beck, P.~G., Bedding, T.~R., Mosser, B., et al. 2011a, Science, 332, 205
\bibitem[]{} Beck, P.~G., Montalb\'an, Kallinger, T., et al.
2011b, Nature, in the press, [arXiv: 1112.2825]
\bibitem[]{} Bedding, T.~R., Mosser, B., \& Huber, D.,  et al. 2011, Nature, 471, 606
\bibitem[]{} Bedding, T.~R., \& Kjeldsen, H.\ 2003, PASA, 20, 203
\bibitem[]{} Christensen-Dalsgaard, J. 2004, Sol.Phys., 220, 137
\bibitem[]{} Christensen-Dalsgaard, J. 2011 Asteroseismology, Canary Islands Winter School of Astrophysics, Volume XXII
(ed. P L. Pall\'e, Cambridge University Press) in the press,[arXiv: 1106.5946v1]
\bibitem[]{} Christensen-Dalsgaard, J. \& Gough, D. O. 2001, MNRAS, 326, 1115
\bibitem[]{} Dupret, M. A., Belkacem, K., Samadi, R.,
  et al. 2009, A\&A, 506, 57
\bibitem[]{} Dziembowski, W. 1971, Acta Astron., 21, 289
\bibitem[]{} Dziembowski, W. 1977, Acta Astron., 27, 95
\bibitem[]{} Dziembowski, W. A., Gough, D. O., Houdek, G,
\& Sienkiewicz, R. 2001, MNRAS, 328, 601
\bibitem[]{} Dziembowski, W. A., Soszy\'nski, I. 2010, A\&A, 524, 88
\bibitem[]{} Hekker, S., Gilliland, R. L., Elsworth, Y., et al. 2011, MNRAS, 414, 2594
\bibitem[]{} Kjeldsen, H. \& Bedding, T. R. 1995, A\&A, 293, 87
\bibitem[]{} Kumar, P. \& Goodman, J. 1996, ApJ, 466, 946
\bibitem[]{} Montalb\'an, J., Miglio, A., Noels, A. Scuflaire, R., Ventura, P. 2010, ApJL, 721, 182
\bibitem[]{} Mosser, B., Belkacem, Goupil, M.-J. Goupil, et al.
2010, A\&A, 517, 22
\bibitem[]{} Mosser, B., Barban, C., Montalb\'an, J. 2011, A\&A,
532, 86
\bibitem[]{} Osaki, Y. 1977, Pub. Astron. Soc. Japan, 29, 234
\bibitem[]{} Pietrinferni, A., Cassisi, S., Salaris, M., \& Castelli,
F. 2006, ApJ, 642, 797
\bibitem[]{} Soszy\'nski, I., Dziembowski, W. A., Udalski, A.,
    et al. 2007,  Acta Astron., 57, 1
\bibitem[]{} Takata, M., 2005, Pub. Astron. Soc. Japan, 57, 275
\bibitem[]{} Takata, M., 2006a, Pub. Astron. Soc. Japan, 58,
759
\bibitem[]{} Takata, M., 2006b, Pub. Astron. Soc. Japan, 58, 893
\end{thebibliography}
\end{document}